\documentclass[12pt,english]{article}
\usepackage[T1]{fontenc}
\usepackage[latin1]{inputenc}
\usepackage{a4}
\usepackage{babel}
\usepackage{graphics}

\makeatletter

\providecommand{\LyX}{L\kern-.1667em\lower.25em\hbox{Y}\kern-.125emX\@}

\makeatother
\begin{document}

\title{Thermal entanglement properties of small spin clusters}

\author{Indrani Bose and Amit Tribedi}

\maketitle
{\centering Department of Physics\par}

{\centering Bose Institute\par}

{\centering 93/1, Acharya Prafulla Chandra Road\par}

{\centering Kolkata - 700 009, India\par}

\begin{abstract}
Exchange interactions in spin systems can give rise to quantum entanglement
in the ground and thermal states of the systems. In this paper, we
consider a spin tetramer, with spins of magnitude \( \frac{1}{2} \),
in which the spins interact via nearest-neighbour, diagonal and four-spin
interactions of strength \( J_{1} \), \( J_{2} \) and \( K \) respectively.
The ground and thermal state entanglement properties of the tetramer
are calculated analytically in the various limiting cases. Both bipartite
and multipartite entanglements are considered and a signature of quantum
phase transition (QPT), in terms of the entanglement ratio, is identified.
The first order QPT is accompanied by discontinuities in the nearest-neighbour
and diagonal concurrences. The magnetic properties of a \( S=\frac{1}{2} \)
AFM polyoxovanadate compound, \( V12 \), are well explained by tetramers,
with \( J_{2}=0 \), \( K \) = 0, in which the spins interact via
the isotropic Heisenberg exchange interaction Hamiltonian. Treating
the magnetic susceptibility \( \chi  \) as an entanglement witness
(EW), an estimate of the lower bound of the critical entanglement
temperature, \( T_{c} \), above which the entanglement between two
individual spins disappears in the experimental compound, is determined.
Two other cases considered include the symmetric tetramer, i.e. tetrahedron
( \( J_{1}=J_{2} \), \( K \)=0 ) and the symmetric trimer. In both
the cases, there is no entanglement between a pair of spins in the
thermal state but multipartite entanglement is present. A second EW
based on energy provides an estimate of the entanglement temperature,
\( T_{E} \), below which the thermal state is definitely entangled.
This EW detects bipartite entanglement in the case of the tetramer
describing a square of spins ( the case of \( V12 \) ) and multipartite
entanglement in the cases of the tetrahedron and the symmetric trimer.
\end{abstract}

\section*{I. Introduction}

Entanglement is a fundamental property of quantum mechanical systems
and gives rise to an excess of correlations in a system over and above
those expected from classical considerations \cite{key-1}. A pure
state is said to be entangled if it does not factorize , i.e., cannot
be written as a product of individual wave functions. A well -known
example of an entangled state is the singlet state of two spin-\( \frac{1}{2} \)
particles, \( \frac{1}{\sqrt{2}}\left( \left| \uparrow \downarrow \right\rangle -\left| \downarrow \uparrow \right\rangle \right)  \),
which cannot be written as a product of the spin states of individual
spins. Measurement on one component of an entangled pair fixes the
state of the other implying non-local correlations. In the case of
a mixed state, entanglement occurs if the density matrix is not a
convex sum of product states. The importance of entanglement derives
from its essential role in applications related to quantum information
and communication. Candidate systems for implementing the application
protocols include spin systems in which exchange interactions give
rise to entanglement \cite{key-2,key-3,key-4,key-5}.

Entanglement is a resource which can be created, manipulated and destroyed.
It can be of different types, e.g., bipartite, multipartite, localizable
\cite{key-6}, zero-temperature, finite-temperature etc. for which
appropriate quantification measures are available. Bipartite (multipartite)
entanglement involves two (more than two) subsystems. The entanglement
between a pair of spins belonging to a chain of interacting spins
provides an example of bipartite entanglement. Bipartite and to a
lesser extent multipartite entanglement properties of a variety of
spin models have been studied so far at both zero and finite temperatures
and including an external magnetic field \cite{key-7,key-8,key-9,key-10,key-11,key-12,key-13,key-14,key-15,key-16}.
These studies show that the amount of entanglement can be changed
by changing the temperature T and/or the external magnetic field.
Since entanglement involves non-local correlations of purely quantum
origin, an issue of considerable interest is whether entanglement
develops special features in the vicinity of a quantum phase transition
(QPT). A QPT occurs at \( T=0 \) and is brought about by tuning some
system parameter, say, the exchange interaction strength or an external
variable like the magnetic field to a critical value \cite{key-17}.
In a QPT, the ground state of the system undergoes qualitative changes
which in turn affects the entanglement properties in the ground state.
Some recent studies have explored the relation between entanglement
and QPT in a variety of spin models and the main conclusion is that
certain entanglement-related quantities exhibit features like scaling
and singularity in the vicinity of a quantum critical point (QCP)
\cite{key-15,key-16,key-18,key-19,key-20,key-21,key-22}. In the case
of first-order QPTs, the ground state concurrences may change discontinuously
at the transition point \cite{key-23,key-24,key-25,key-26}.

\noindent The influence of quantum criticality extends also to finite
temperatures so that measurements of appropriate observables provide
signatures of QPT. At finite \( T \), the system in thermal equilibrium
is described by the density operator, \( \rho \left( T\right) =\frac{1}{Z}exp\left( -\frac{H}{k_{B}T}\right)  \),
where \( H \) is the Hamiltonian, \( Z \) the partition function
and \( k_{B} \) the Boltzmann constant. A thermal state remains entangled
upto a critical temperature \-\( T_{c} \) beyond which the state
becomes separable, i.e., the entanglement falls to zero. Experimental
demonstrations of entanglement are mostly confined to the microworld,
i.e., to systems consisting of a few photons, atoms or ions. There
is now experimental evidence that entanglement can also affect the
macroscopic properties of solids. This has been shown in the insulating
magnetic compound \( LiHo_{x}Y_{1-x}F_{4} \) the specific heat and
the susceptibility data of which can only be explained if quantum
entanglement of the relevant states is explicitly taken into account
\cite{key-27,key-28}. Measures of thermal entanglement based on the
thermal density matrix require a knowledge of both the eigenvalues
and the eigenvectors of \( H \). On the other hand, there are suggestions
that macroscopic thermodynamic observables can serve as entanglement
witnesses so that a measurement of these quantities can provide the
evidence for entanglement \cite{key-11,key-28,key-29,key-30,key-31}.
An entanglement witness (EW) is an observable the expectation value
of which is positive in unentangled, i.e., separable states and negative
in entangled states \cite{key-32,key-33,key-34}. The thermodynamic
observables which have been proposed as EWs include internal energy
and magnetization and magnetic susceptibility \cite{key-28,key-31}.
The latter has been used as an EW in the spin-\( \frac{1}{2} \) alternating
bond antiferromagnet \( Cu(NO_{3})_{2}2.5D_{2}O(CN) \). The compound
can be considered as a chain of uncoupled spin dimers since the ratio
of the inter-dimer to the intra-dimer exchange interaction strengths
is approximately \( 0.24 \), i.e., low. For separable (unentangled)
states, the magnetic susceptibility obeys the inequality 

\begin{equation}
\label{1}
\chi \geq \frac{(g\mu _{B})^{2}N}{k_{B}T}\frac{1}{6}
\end{equation}
 where \( g \) is the Land\'{e} splitting factor, \( \mu _{B} \)
the Bohr magneton and \( N \) the number of spins in the system.
Entanglement is present in the system if the inequality in \( (1) \)
is violated. The intersection point of the curve representing the
EW (equality in \( (1) \) ) and the experimental \( \chi  \) versus
\( T \) curve defines the critical temperature \( T_{c} \) below
which entanglement is present in the system. The experimental estimate
of \( T_{c}\simeq 5K \) is in good agreement with the theoretical
value of the critical temperature at which the pairwise thermal entanglement
(entanglement between two spins ), as measured by the concurrence,
falls to zero.

Determination of the entanglement properties of an interacting spin
system is a theoretical challenge as the eigenstates and eigenvalues
are not known exactly when the number of spins is large. Most of the
calculations are confined to systems containing a few spins so that
exact diagonalization is possible. Studies on finite quantum spin
systems acquire significant relevance in the context of molecular
or nanomagnets. In such magnetic systems, the dominant exchange interactions
are often confined to small spin clusters. The inter-cluster exchange
interactions are much weaker in comparison so that the compounds can
be assumed to consist of independent spin clusters. A recent study
provides a number of examples of molecular magnets the thermodynamic
and neutron scattering properties of which can be well described by
small spin clusters like dimers, trimers and tetramers \cite{key-35}.
As in Ref. \cite{key-28}, one can study the entanglement properties
of the molecular magnets by treating the susceptibility \( \chi  \)
as an EW. The earlier work dealt with spin dimers for which only pairwise
entanglement is possible. In this paper, we consider clusters of three
(trimer) and four (tetramer) spins in which pairwise entanglement
between individual spins does not exhaust the total entanglement.
The tetramer Hamiltonian contains both bilinear and four-spin interactions.
The ground state and thermal entanglement properties of the tetramers
are determined analytically. The influence of multispin interactions
on entanglement is further determined. The system exhibits QPTs at
special values of the exchange interaction strengths. A signature
of the QPT via the so-called entanglement ratio is identified. A distinct
signature of first order QPT is provided by jumps in the amounts of
entanglement associated with n.n. and diagonal spin pairs. The magnetic
properties of the polyoxovanadate compound, \( \left( NHEt\right) _{3}\left[ V^{IV}_{8}V_{4}^{V}As_{8}O_{40}\left( H_{2}O\right) \right] .H_{2}O \)
(designated as \( V12 \) ) are well explained by spin\( -\frac{1}{2} \)
AFM tetramers, with only nearest-neighbour (n.n.) interactions, and
described by the isotropic Heisenberg exchange interaction Hamiltonian
\cite{key-35,key-36}. The experimental data on the magnetic susceptibility
of this compound are available. Treating \( \chi  \) as an EW, the
critical entanglement temperature, \( T_{c} \), below which entanglement
is certainly present in the system, is determined. The cases of the
\( S=\frac{1}{2} \) AFM symmetric trimer and tetrahedron are also
considered.

Dowling et al.\cite{key-33} have introduced the concept of the entanglement
gap, defined to be the difference in the energies of the minimum energy,
\( E_{sep} \), that a separable state may attain and the ground state
energy \( E_{0} \). If the energy of the system falls within the
entanglement gap, the state of the system is entangled. The entanglement
gap temperature, \( T_{E} \), is defined to be the temperature at
which the thermal energy \( U(T_{E})=E_{sep} \), the minimum separable
energy. Below \( T_{E} \), the thermal state of the system is bound
to be entangled. We obtain an estimate of \( T_{E} \) in the cases
of a single square of spins (the case of \( V12 \)), a tetrahedron
and a symmetric trimer. In the last two cases the critical entanglement
temperature \( T_{c} \), determined by using \( \chi  \) as an EW,
is identical to the entanglement gap temperature \( T_{E} \).

\begin{figure}
{\centering \includegraphics{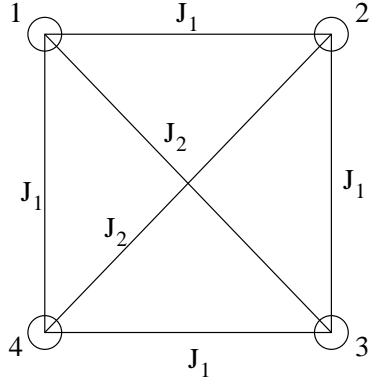} \par}

\caption{A tetramer of spins of magnitude \protect\( \frac{1}{2}\protect \).
\protect\( J_{1}\protect \) and \protect\( J_{2}\protect \) denote
the strengths of the n.n. and diagonal exchange interactions. The
four-spin interactions are not shown.}
\end{figure}

\section*{II. Entanglement properties of \protect\( S=\frac{1}{2}\protect \)
AFM tetramer}

We consider a tetramer of spins of magnitude \( \frac{1}{2} \) (Fig.
1) described by the AFM Heisenberg exchange interaction Hamiltonian\begin{equation}
\label{2}
\begin{array}{ccc}
H & = & J_{1}\left( S_{1}.S_{2}+S_{2}.S_{3}+S_{3}.S_{4}+S_{4}.S_{1}\right) +J_{2}(S_{1}.S_{3}+S_{2}.S_{4})\\
 &  & +K_{1}(S_{1}.S_{2})(S_{3}.S_{4})+K_{1}(S_{2}.S_{3})(S_{1}.S_{4})+K_{2}(S_{1}.S_{3})(S_{2}.S_{4})
\end{array}
\end{equation}
where \( S_{i} \) is the spin operator at the \( ith \) site of
the square plaquette, \( J_{1} \) is the strength of the n.n. exchange
interaction, \( J_{2} \) that of the diagonal exchange interaction
and \( K_{1} \), \( K_{2} \) are the strengths of the four-spin
exchange interactions. The entanglement properties of the four-spin
cluster have earlier been studied analytically only for \( J_{1}\neq 0 \)
\cite{key-37} and numerically for both \( J_{1}\neq 0,J_{2}\neq 0 \)
\cite{key-38}. We now determine the entanglement properties analytically
for the general case in Eq. (2). The \( z \)-component of the total
spin, \( S^{tot}_{z} \), is a conserved quantity so that the eigenvalue
problem can be solved in the separate subspaces corresponding to the
different values of \( S^{tot}_{z} \). The results are displayed
in the following (\( E_{i} \) , \( i=1,....,16 \), is the energy
eigenvalue):

\( S^{tot}_{z}=+2 \)

\begin{equation}
\label{3}
\begin{array}{c}
\psi _{1}=\left| \uparrow \uparrow \uparrow \uparrow \right\rangle \\
E_{1}=\left( J_{1}+\frac{J_{2}}{2}+\frac{K_{1}}{8}+\frac{K_{2}}{16}\right) 
\end{array}
\end{equation}

\( S^{tot}_{z}+1 \)

\begin{equation}
\label{4}
\begin{array}{c}
\psi _{2}=\frac{1}{\sqrt{2}}\left( \left| \uparrow \uparrow \uparrow \downarrow \right\rangle -\left| \uparrow \downarrow \uparrow \uparrow \right\rangle \right) \\
E_{2}=-\left( \frac{J_{2}}{2}+\frac{K_{1}}{8}+\frac{3K_{2}}{16}\right) 
\end{array}
\end{equation}

\begin{equation}
\label{5}
\begin{array}{c}
\psi _{3}=\frac{1}{\sqrt{2}}\left( \left| \uparrow \uparrow \downarrow \uparrow \right\rangle -\left| \downarrow \uparrow \uparrow \uparrow \right\rangle \right) \\
E_{3}=-\left( \frac{J_{2}}{2}+\frac{K_{1}}{8}+\frac{3K_{2}}{16}\right) 
\end{array}
\end{equation}

\begin{equation}
\label{6}
\begin{array}{c}
\psi _{4}=\frac{1}{\sqrt{4}}\left( \left| \uparrow \uparrow \uparrow \downarrow \right\rangle +\left| \uparrow \uparrow \downarrow \uparrow \right\rangle +\left| \uparrow \downarrow \uparrow \uparrow \right\rangle +\left| \downarrow \uparrow \uparrow \uparrow \right\rangle \right) \\
E_{4}=\left( J_{1}+\frac{J_{2}}{2}+\frac{K_{1}}{8}+\frac{K_{2}}{16}\right) 
\end{array}
\end{equation}

\begin{equation}
\label{7}
\begin{array}{c}
\psi _{5}=\frac{1}{\sqrt{4}}\left( \left| \uparrow \uparrow \uparrow \downarrow \right\rangle +\left| \uparrow \downarrow \uparrow \uparrow \right\rangle -\left| \uparrow \uparrow \downarrow \uparrow \right\rangle -\left| \downarrow \uparrow \uparrow \uparrow \right\rangle \right) \\
E_{5}=\left( -J_{1}+\frac{J_{2}}{2}-\frac{3K_{1}}{8}+\frac{K_{2}}{16}\right) 
\end{array}
\end{equation}

\( S^{tot}_{z}=0 \)

\begin{equation}
\label{8}
\begin{array}{c}
\psi _{6}=\frac{1}{\sqrt{2}}\left( \left| \uparrow \uparrow \downarrow \downarrow \right\rangle -\left| \downarrow \downarrow \uparrow \uparrow \right\rangle \right) \\
E_{6}=-\left( \frac{J_{2}}{2}+\frac{K_{1}}{8}+\frac{3K_{2}}{16}\right) 
\end{array}
\end{equation}

\begin{equation}
\label{9}
\begin{array}{c}
\psi _{7}=\frac{1}{\sqrt{2}}\left( \left| \uparrow \downarrow \downarrow \uparrow \right\rangle -\left| \downarrow \uparrow \uparrow \downarrow \right\rangle \right) \\
E_{7}=-\left( \frac{J_{2}}{2}+\frac{K_{1}}{8}+\frac{3K_{2}}{16}\right) 
\end{array}
\end{equation}

\begin{equation}
\label{10}
\begin{array}{c}
\psi _{8}=\frac{1}{\sqrt{2}}\left( \left| \uparrow \downarrow \uparrow \downarrow \right\rangle -\left| \downarrow \uparrow \downarrow \uparrow \right\rangle \right) \\
E_{8}=\left( -J_{1}+\frac{J_{2}}{2}-\frac{3K_{1}}{8}+\frac{K_{2}}{16}\right) 
\end{array}
\end{equation}

\begin{equation}
\label{11}
\begin{array}{c}
\psi _{9}=\frac{1}{\sqrt{6}}\left( \left| \uparrow \uparrow \downarrow \downarrow \right\rangle +\left| \uparrow \downarrow \downarrow \uparrow \right\rangle +\left| \downarrow \downarrow \uparrow \uparrow \right\rangle +\left| \downarrow \uparrow \uparrow \downarrow \right\rangle +\left| \uparrow \downarrow \uparrow \downarrow \right\rangle +\left| \downarrow \uparrow \downarrow \uparrow \right\rangle \right) \\
E_{9}=\left( J_{1}+\frac{J_{2}}{2}+\frac{K_{1}}{8}+\frac{K_{2}}{16}\right) 
\end{array}
\end{equation}

\begin{equation}
\label{12}
\begin{array}{c}
\psi _{10}=\frac{1}{\sqrt{4}}\left( \left| \uparrow \uparrow \downarrow \downarrow \right\rangle +\left| \downarrow \downarrow \uparrow \uparrow \right\rangle -\left| \uparrow \downarrow \downarrow \uparrow \right\rangle -\left| \downarrow \uparrow \uparrow \downarrow \right\rangle \right) \\
E_{10}=\left( -\frac{3J_{2}}{2}+\frac{3K_{1}}{8}+\frac{9K_{2}}{16}\right) 
\end{array}
\end{equation}

\begin{equation}
\label{13}
\begin{array}{c}
\psi _{11}=\frac{1}{\sqrt{12}}\left( 2\left| \uparrow \downarrow \uparrow \downarrow \right\rangle +2\left| \downarrow \uparrow \downarrow \uparrow \right\rangle -\left| \uparrow \uparrow \downarrow \downarrow \right\rangle -\left| \uparrow \downarrow \downarrow \uparrow \right\rangle -\left| \downarrow \downarrow \uparrow \uparrow \right\rangle -\left| \downarrow \uparrow \uparrow \downarrow \right\rangle \right) \\
E_{11}=\left( -2J_{1}+\frac{J_{2}}{2}+\frac{7K_{1}}{8}+\frac{K_{2}}{16}\right) 
\end{array}
\end{equation}

\( S^{tot}_{z}=-1 \)

\begin{equation}
\label{14}
\begin{array}{c}
\psi _{12}=\frac{1}{\sqrt{2}}\left( \left| \downarrow \downarrow \downarrow \uparrow \right\rangle -\left| \downarrow \uparrow \downarrow \downarrow \right\rangle \right) \\
E_{12}=-\left( \frac{J_{2}}{2}+\frac{K_{1}}{8}+\frac{3K_{2}}{16}\right) 
\end{array}
\end{equation}

\begin{equation}
\label{15}
\begin{array}{c}
\psi _{13}=\frac{1}{\sqrt{2}}\left( \left| \downarrow \downarrow \uparrow \downarrow \right\rangle -\left| \uparrow \downarrow \downarrow \downarrow \right\rangle \right) \\
E_{13}=-(\frac{J_{2}}{2}+\frac{K_{1}}{8}+\frac{3K_{2}}{16})
\end{array}
\end{equation}

\begin{equation}
\label{16}
\begin{array}{c}
\psi _{14}=\frac{1}{\sqrt{4}}\left( \left| \downarrow \downarrow \downarrow \uparrow \right\rangle +\left| \downarrow \downarrow \uparrow \downarrow \right\rangle +\left| \downarrow \uparrow \downarrow \downarrow \right\rangle +\left| \uparrow \downarrow \downarrow \downarrow \right\rangle \right) \\
E_{14}=\left( J_{1}+\frac{J_{2}}{2}+\frac{K_{1}}{8}+\frac{_{K_{2}}}{16}\right) 
\end{array}
\end{equation}

\begin{equation}
\label{17}
\begin{array}{c}
\psi _{15}=\frac{1}{\sqrt{4}}\left( \left| \downarrow \downarrow \downarrow \uparrow \right\rangle +\left| \downarrow \uparrow \downarrow \downarrow \right\rangle -\left| \downarrow \downarrow \uparrow \downarrow \right\rangle -\left| \uparrow \downarrow \downarrow \downarrow \right\rangle \right) \\
E_{15}=\left( -J_{1}+\frac{J_{2}}{2}-\frac{3K_{1}}{8}+\frac{K_{2}}{16}\right) 
\end{array}
\end{equation}

\( S^{tot}_{z}=-2 \)

\begin{equation}
\label{18}
\begin{array}{c}
\psi _{16}=\left| \downarrow \downarrow \downarrow \downarrow \right\rangle \\
E_{16}=\left( J_{1}+\frac{J_{2}}{2}+\frac{K_{1}}{8}+\frac{K_{2}}{16}\right) 
\end{array}
\end{equation}

We first discuss the ground state \( (T=0) \) entanglement properties.
There are five distinct eigenvalues:

\begin{equation}
\label{19}
\begin{array}{c}
e_{1}=E_{1}=E_{4}=E_{9}=E_{14}=E_{16}=(J_{1}+\frac{J_{2}}{2}+\frac{K_{1}}{8}+\frac{K_{2}}{16})\\
e_{2}=E_{2}=E_{3}=E_{6}=E_{7}=E_{12}=E_{13}=-\left( \frac{J_{2}}{2}+\frac{K_{1}}{8}+\frac{3K_{2}}{16}\right) \\
e_{3}=E_{5}=E_{8}=E_{15}=\left( -J_{1}+\frac{J_{2}}{2}-\frac{3K_{1}}{8}+\frac{K_{2}}{16}\right) \\
e_{4}=E_{10}=\left( -\frac{3J_{2}}{2}+\frac{3K_{1}}{8}+\frac{9K_{2}}{16}\right) \\
e_{5}=E_{11}=\left( -2J_{1}+\frac{J_{2}}{2}+\frac{7K_{1}}{8}+\frac{K_{2}}{16}\right) 
\end{array}
\end{equation}
For simplicity, let us put \( K_{1}=K_{2}=K \). When \( J_{2}<J_{1} \)
and \( K<\frac{4J_{1}}{5} \), the ground state is non-degenerate
with eigenvalue \( e_{5} \). When \( J_{2}<J_{1} \) and \( K>\frac{4J_{1}}{5} \),
the ground state is three-fold degenerate with eigenvalue \( e_{3} \).
Thus \( K=\frac{4J_{1}}{5} \) \( (J_{2}<J_{1}) \) is a QCP. When
\( J_{1}<J_{2} \) and \( K<\frac{4J_{2}}{5} \), the ground state
is non-degenerate with eigenvalue \( e_{4} \). When \( J_{1}<J_{2} \)
and \( K>\frac{4J_{2}}{5} \), the ground state is six-fold degenerate
with eigenvalue \( e_{2} \) . In this case a QPT occurs at \( K=\frac{4J_{2}}{5} \).
With \( K< \) \( \frac{4J_{1}}{5} \), a QPT occurs at \( J_{1}=J_{2} \)
when the ground state changes from \( \psi _{11} \) to \( \psi _{10}. \)
In this paper, we focus our attention on this last QPT. The states
\( \psi _{11} \) and \( \psi _{10} \) describe two resonating valence
bond (RVB) states, \( \psi _{RVB1} \) and \( \psi _{RVB2} \) respectively.
Figure 2 gives a pictorial representation of \( \psi _{RVB1} \) and
\( \psi _{RVB2} \). The solid lines represent singlets (valence bonds)
and the arrow signs follow the phase convention that a VB between
the sites \( i \) and \( j \) represents the spin configuration
\( \frac{1}{\sqrt{2}}\left[ \left| \uparrow (i)\downarrow (j)\right\rangle -\left| \downarrow (i)\uparrow (j)\right\rangle \right] , \)
if the arrow points away from the site \( i \).

\begin{figure}
{\centering \includegraphics{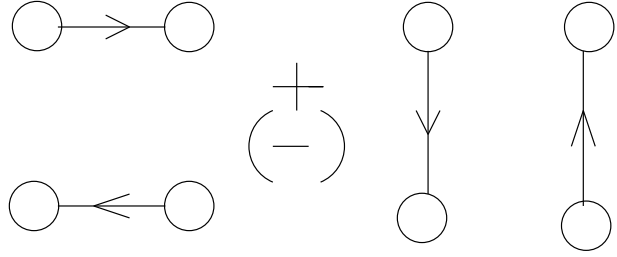} \par}

\caption{The two resonating valence bond (RVB) states, \protect\( \psi _{RVB1}\protect \)(
+ sign) and \protect\( \psi _{RVB2}\protect \)( - sign). A solid
line represents a singlet spin configuration. The arrow convention
is explained in the text}
\end{figure}

A measure of entanglement between the spins at sites \( i \) and
\( j \) is given by concurrence \cite{key-7,key-8}. To calculate
this, a knowledge of the reduced density matrix \( \rho _{ij} \)
is required. This is obtained from the ground state wave function
by tracing out all the spin degrees of freedom except those of the
spins at the sites \( i \) and \( j \). Let \( \rho _{ij} \) be
defined as a matrix in the standard basis \( \left\{ \left| \uparrow \uparrow \right\rangle ,\left| \uparrow \downarrow \right\rangle ,\left| \downarrow \uparrow \right\rangle ,\left| \downarrow \downarrow \right\rangle \right\} . \)
One can define the spin-reversed density matrix as \( \widetilde{\rho }=\left( \sigma _{y}\otimes \sigma _{y}\right) \rho ^{*}\left( \sigma _{y}\otimes \sigma _{y}\right)  \),
where \( \sigma _{y} \) is the Pauli matrix. The concurrence \( C \)
is given by \( C=max\left\{ \lambda _{1}-\lambda _{2}-\lambda _{3}-\lambda _{4},0\right\}  \),
where the \( \lambda _{i} \)'s are the square roots of the eigenvalues
of the matrix \( \rho \widetilde{\rho } \) in descending order. \( C=0 \)
implies an unentangled state whereas \( C=1 \) corresponds to maximum
entanglement. The reduced density matrix in the standard basis has
the structure 

\begin{equation}
\label{20}
\rho _{ij}=\left( \begin{array}{cccc}
u & 0 & 0 & 0\\
0 & \omega _{1} & y^{*} & 0\\
0 & y & \omega _{2} & 0\\
0 & 0 & 0 & v
\end{array}\right) 
\end{equation}
and the concurrence \( C_{ij} \) has the simple form 

\begin{equation}
\label{21}
C_{ij}=2\textrm{ }max\left( 0,\left| y\right| -\sqrt{uv}\right) 
\end{equation}
If the ground state is degenerate, the \( T=0 \) ensemble is described
by a density matrix which is an equal mixture of contributions from
all possible ground states. The density matrix is a limiting case
of the thermal density matrix as \( T\rightarrow 0 \). The state
\( \psi _{RVB1} \) is the ground state for \( J_{2}<J_{1} \) and
\( K<\frac{4J_{1}}{5}. \) In this case, the n.n. concurrences \( C_{12}=C_{23}=C_{34}=C_{41}=0.5 \),
i.e., the n.n. spin pairs are entangled in equal amounts. The magnitude
of the concurrence is independent of \( J_{1},J_{2} \) and \( K \)
as long as \( \psi _{RVB1} \) remains the ground state. The concurrences
\( C_{13} \) and \( C_{24} \) are zero, i.e., the spins at the ends
of a diagonal are unentangled. At the QCP, \( J_{1}=J_{2}=J \) (\( K<\frac{4J}{5} \)),
the concurrences \( C_{12},C_{23},C_{34},C_{41} \) and \( C_{13},C_{24} \)
are all equal to zero. The ground state at this point is doubly degenerate
with wave functions \( \psi _{RVB1} \) and \( \psi _{RVB2} \). For
\( J_{2}>J_{1} \) and \( K<\frac{4J_{2}}{5} \), the ground state
is given by \( \psi _{RVB2} \). The n.n. concurrences \( C_{12},C_{23},C_{34} \)
and \( C_{41} \) are now zero whereas \( C_{13}=C_{24}=1 \). The
spin configuration described by \( \psi _{RVB2} \) (Fig. 2) can alternatively
be described as consisting of VBs, i.e., singlets across the diagonals.
Since a singlet is maximally entangled, \( C_{13}=C_{24}=1 \). The
entanglement properties of a system can further be analyzed in terms
of a quantity known as the one-tangle \( \tau _{1} \) which is a
measure of the entanglement between a spin and the remainder of the
system \cite{key-39,key-40,key-48}. It is given as \( \tau _{1}=4det\rho ^{(1)} \)
where \( \rho ^{(1)} \) is the single-site reduced density matrix.
In a translationally invariant system, \( \tau _{1} \) provides a
global estimate of the entanglement present whereas the concurrence
gives a measure of the pairwise entanglement between two individual
spins. When \( \tau _{1}=0 \), there is no entanglement in the ground
state, i.e., the state becomes separable. The Coffman-Kundu-Wootters
(CKW) conjecture \cite{key-39}, originally proposed for a three-qubit
system, can be generalized to yield the inequality

\begin{equation}
\label{22}
\tau _{1}\geq \tau _{2}=\sum _{j\neq i}C^{2}_{ij}
\end{equation}
where \( \tau _{1} \) represents the one-tangle corresponding to
the entanglement between the \( ith \) qubit (spin) and the rest
of the system and \( C^{2}_{ij} \) is the square of the concurrence
associated with the pairwise entanglement between the \( ith \) and
\( jth \) qubits. The inequality in (22) shows that the pairwise
entanglement is not the sole entanglement in the system. For the four-spin
cluster, \( \tau  \)\( _{1} \) has the value \( 1 \) when \( \psi _{RVB1} \)
and/or \( \psi _{RVB2} \) are the ground states. The ratio \( R=\frac{\tau _{2}}{\tau _{1}} \)
quantifies the relative contribution of the pairwise entanglement
and has values \( \frac{1}{2} \) and \( 1 \) in the ground states
\( \psi _{RVB1} \) and \( \psi _{RVB2} \) respectively. Roscilde
et al. \cite{key-40,key-48} have shown that the value of \( R \)
reaches a minimum (not zero) at the QCP of \( S=\frac{1}{2} \) XYX
AFMs in an external magnetic field. In the present case, we have a
first order QPT. At the transition point J\( _{1} \)=J\( _{2} \)=J
(\( K<\frac{4J}{5} \)), the ground state is doubly degenerate so
that the system is in a mixed state. The entanglement measure \( \tau _{1} \),
 defined for pure states, needs to be generalized to the case of mixed
states. This is done \cite{key-39} by considering all possible pure
state decompositions of the density matrix \( \rho  \). For each
of the decompositions, one can determine the average value of \( \tau _{1} \).
The minimum of the average over all decompositions is taken to be
\( \tau ^{min}_{1} \) which replaces \( \tau _{1} \) in the CKW
inequality in Eq. (22). While calculation of \( \tau ^{min}_{1} \)
is difficult, one can readily see that \( R \) (\( R=\frac{\tau _{2}}{\tau _{1}^{min}} \))
at the QPT point either has the value zero (\( \tau  \)\( _{2} \)=0,
\( \tau ^{min}_{1} \)\( \neq  \)0) or is undefined (\( \tau  \)\( _{2} \)=0,
\( \tau ^{min}_{1} \)\( = \)0) . In the former case, the value of
\( R \) reaches a minimum at the transition point. In both the cases
\( R \) has distinct values on both sides of the transition point.
In \( \psi _{RVB1} \), two-spin entanglements exhaust the one-tangle
whereas the opposite is true in the case of \( \psi _{RVB2} \). A
clearer signature of first order QPT is provided by the jumps in both
the n.n. and diagonal concurrences \cite{key-18,key-23}. In the present
model, the n.n. concurrences \( C_{12} \), \( C_{23} \), \( C_{34} \)
and \( C_{14} \) are equal to 0.5 in the ground state \( \psi _{RVB1} \)
and zero at the transition point as well as in the ground state \( \psi _{RVB2} \).
The diagonal concurrences \( C_{13} \) and \( C_{24} \) are equal
to 1 in \( \psi _{RVB2} \) and zero at the transition point as well
as in the state \( \psi _{RVB1} \). The jumps in the magnitudes of
the concurrences are associated with the jumps in the density matrix
elements, a typical feature of first order QPTs \cite{key-18}.

We now discuss the finite temperature entanglement properties of the
spin tetramer. The thermal density matrix, \( \rho (T)=\frac{1}{Z}exp(-\beta H) \)
\( (\beta =\frac{1}{k_{B}T}) \), now replaces the ground state density
matrix with Z denoting the partition function of the system. The reduced
thermal density matrix \( \rho _{ij}(T) \) has the same form as in
(20) with \( C_{ij}(T) \) given by

\begin{equation}
\label{23}
C_{ij}(T)=\frac{2}{Z}max\left( 0,\left| y(T)\right| -\sqrt{u(T)v(T)}\right) 
\end{equation}
For the four-spin cluster, the thermal density matrix is 

\begin{equation}
\label{24}
\rho (T)=\frac{1}{Z}\sum ^{16}_{k=1}exp(-\beta E_{k})\left| \psi _{k}\right\rangle \left\langle \psi _{k}\right| 
\end{equation}
where the \( \left| \psi _{k}\right\rangle  \)\( 's \) and the \( E_{k} \)\( 's \)
are given in equations (3)-(18). The matrix elements \( u,v \) and
\( y \) of the reduced thermal density matrix \( \rho _{12}(T) \)
are

\begin{equation}
\label{25}
\begin{array}{c}
u=v=\frac{5}{3}e^{-\beta e_{1}}+\frac{3}{2}e^{-\beta e_{2}}+\frac{1}{2}e^{-\beta e_{3}}+\frac{1}{4}e^{-\beta e_{4}}+\frac{1}{12}e^{-\beta e_{5}}\\
y=\frac{5}{6}e^{-\beta e_{1}}-\frac{1}{2}e^{-\beta e_{3}}-\frac{1}{3}e^{-\beta e_{5}}
\end{array}
\end{equation}
where the eigenvalues \( e_{i} \)\( 's \) \( (i=1,2,...,5) \) are
given in Eq. (19). Due to translational invariance, the reduced density
matrices for the other n.n. spin pairs have the same matrix elements
as in the case of \( \rho _{12}(T) \). Figure 3 shows \( C_{12} \)
as a function of \( \frac{k_{B}T}{J_{1}} \) for \( \frac{J_{2}}{J_{1}}=0.5 \)
and for \( r=\frac{K}{J_{1}} \) (\( K_{1}=K_{2}=K \)) \( =0.4 \)
\( (a) \), \( 0.2 \) \( (b) \) and \( 0.0 \) \( (c) \). Increase
in the strength of the four-spin interaction reduces the magnitude
of the n.n. concurrence. The value of the concurrence is non-zero
provided \( \left| y\right| -\sqrt{uv} \) (Eq.(23)) is \( >0 \).
One can define a critical temperature \( T_{c} \) beyond which the
entanglement between n.n. spins disappears \cite{key-37,key-43}.
One can show that in the parameter regime of interest, the thermal
entanglement between the diagonal spins is zero so that \( T_{c} \)
can be taken as the critical temperature beyond which the entanglement
between any two spins is zero. The critical temperature \( T_{c} \)
is obtained from \( \left| y\right| -\sqrt{uv}=0 \) (Eq. (23)), i.e.,
as a solution of the equation \begin{equation}
\label{26}
z^{3-\frac{3r_{2}}{4}}-6z^{1+r_{1}+\frac{r_{2}}{2}}-z^{1+2r_{1}-\frac{3r_{2}}{4}}-10=0
\end{equation}
 where \( z=e^{\frac{J_{1}}{k_{B}T}},r_{1}=\frac{J_{2}}{J_{1}} \)
and \( r_{2}=\frac{K}{J_{1}} \). Figure 4 shows a plot of \( \frac{k_{B}T_{c}}{J_{1}} \)
versus \( \frac{J_{2}}{J_{1}} \) for \( r=0.4 \) \( (a) \), \( 0.2 \)
\( (b) \) and \( 0.0 \) \( (c) \). For a fixed value of \( \frac{J_{2}}{J_{1}} \),
the critical entanglement temperature \( T_{c} \) decreases as the
strength of the four-spin interaction increases. \( T_{c} \) tends
to zero as \( \frac{J_{2}}{J_{1}} \) approaches the QCP \( \frac{J_{2}}{J_{1}}=1 \).
For \( J_{2}>J_{1} \) (with \( K<\frac{4J_{2}}{5} \)), the n.n.
concurrences are zero.

\begin{figure}
{\centering \resizebox*{4in}{2in}{\includegraphics{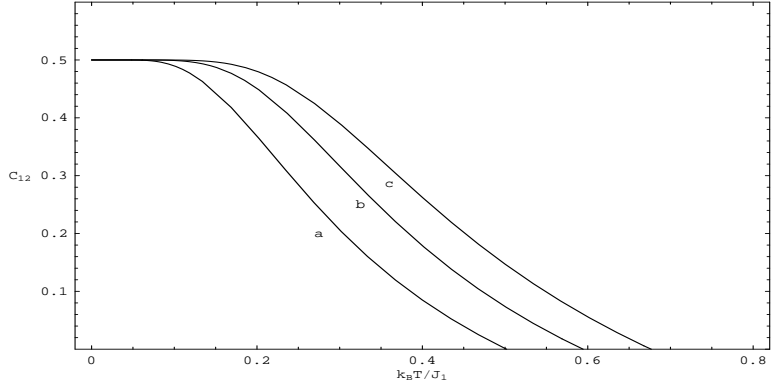}} \par}

\caption{Concurrence \protect\( C_{12}\protect \) as a function of \protect\( \frac{k_{B}T}{J_{1}}\protect \)
for \protect\( \frac{J_{2}}{J_{1}}=0.5\protect \) and for \protect\( r=\frac{K}{J_{1}}(K_{1}=K_{2}=K)\protect \)
\protect\( =0.4\protect \) \protect\( (a)\protect \), \protect\( 0.2\protect \)
\protect\( (b)\protect \) and \protect\( 0.0\protect \) \protect\( (c)\protect \).}
\end{figure}

\begin{figure}
{\centering \resizebox*{4in}{2in}{\includegraphics{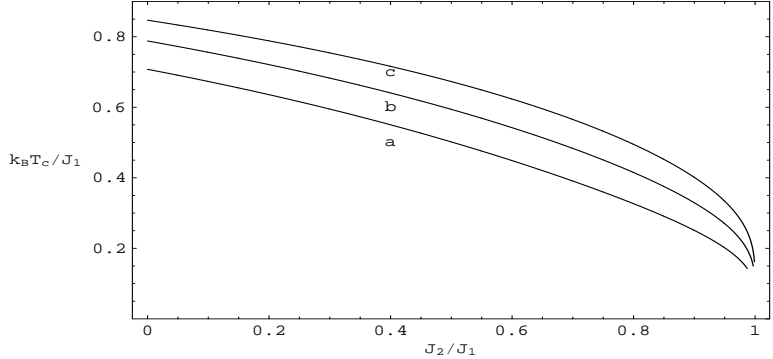}} \par}

\caption{A plot of \protect\( \frac{k_{B}T_{c}}{J_{1}}\protect \), where
\protect\( T_{c}\protect \) is the critical entanglement temperature,
versus \protect\( \frac{J_{2}}{J_{1}}\protect \), for \protect\( r=0.4\protect \)
\protect\( (a)\protect \), \protect\( 0.2\protect \) \protect\( (b)\protect \)
and \protect\( 0.0\protect \) \protect\( (c)\protect \)}
\end{figure}

We next calculate the concurrence for pairwise entanglement between
the spins located at the ends of a diagonal. The matrix elements \( u,v \)
and \( y \) of the reduced thermal density matrix \( \rho _{13}(T) \)
are given by

\begin{equation}
\label{27}
\begin{array}{c}
u=v=\frac{5}{3}e^{-\beta e_{1}}+e^{-\beta e_{2}}+e^{-\beta e_{3}}+\frac{1}{3}e^{-\beta e_{5}}\\
y=\frac{5}{6}e^{-\beta e_{1}}-e^{-\beta e_{2}}+\frac{1}{2}e^{-\beta e_{3}}-\frac{1}{2}e^{-\beta e_{4}}+\frac{1}{6}e^{-\beta e_{5}}
\end{array}
\end{equation}
where the eigenvalues \( e_{i} \)\( 's \) are given in Eq. (19).
The reduced density matrix \( \rho _{24}(T) \) has the same elements
as in the case of \( \rho _{13}(T) \). Figure 5 shows \( C_{13} \)
as a function of \( \frac{k_{B}T}{J_{2}} \) for \( \frac{J_{1}}{J_{2}}=0.5 \)
and for \( r=\frac{K}{J_{2}}(K_{1}=K_{2}=K)=0.4 \) \( (a), \) \( 0.2 \)
\( (b) \) and \( 0.0 \) \( (c) \). Again, at a fixed value of \( \frac{J_{1}}{J_{2}} \),
the magnitude of \( C_{13} \) decreases as the strength of the four-spin
interaction increases. The critical entanglement temperature \( T_{c} \),
beyond which the entanglement between spins located at the ends of
a diagonal disappears, is also the temperature beyond which the pairwise
entanglement between any two spins vanishes since in the parameter
regime of interest the n.n. concurrences are zero at all \( T \).
The critical temperature \( T_{c} \) is obtained as a solution of
the equation

\begin{equation}
\label{28}
z^{2+r_{1}-\frac{3r_{2}}{4}}-3z^{2r_{1}+\frac{r_{2}}{2}}-z^{3r_{1}-\frac{3r_{2}}{4}}-5=0
\end{equation}
where \( z=e^{\frac{J_{2}}{k_{B}T}},r_{1}=\frac{J_{1}}{J_{2}} \)
and \( r_{2}=\frac{K}{J_{2}} \). Figure 6 shows a plot of \( \frac{k_{B}T_{c}}{J_{2}} \)
versus \( \frac{J_{1}}{J_{2}} \) for \( r=0.4 \) \( (a) \), \( 0.2 \)
\( (b) \) and \( 0.0 \) \( (c) \). For a fixed value of \( \frac{J_{1}}{J_{2}} \),
the critical entanglement temperature \( T_{c} \) decreases as the
strength of the four-spin interaction \( K \) increases. \( T_{c} \)
approaches zero as \( \frac{J_{1}}{J_{2}} \) approaches the QCP \( \frac{J_{1}}{J_{2}}=1 \).
The major conclusion one arrives at from an examination of Figs. (3)-(6),
is that, as in the \( T=0 \) case, the two sets of concurrences \( (i) \)
\( C_{12},C_{23},C_{34},C_{41} \) and \( (ii) \) \( C_{13,},C_{24} \)
are mutually exclusive. For finite values of the concurrences belonging
to the first set, the values of the concurrences belonging to the
second set are zero and vice versa.

\begin{figure}
{\centering \resizebox*{4in}{2in}{\includegraphics{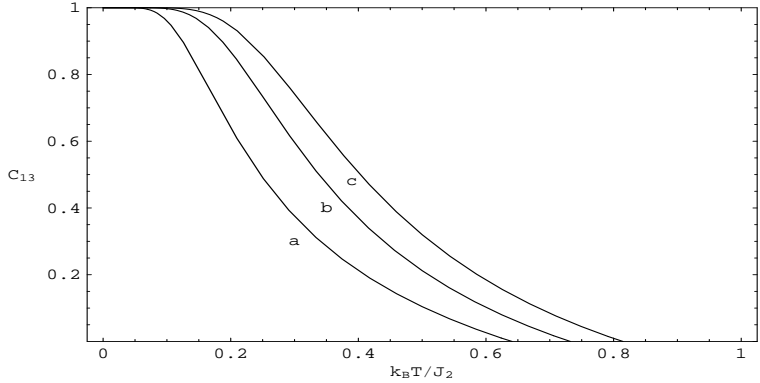}} \par}

\caption{Concurrence \protect\( C_{13}\protect \) as a function of \protect\( \frac{k_{B}T}{J_{2}}\protect \)
for \protect\( \frac{J_{2}}{J_{1}}=0.5\protect \) and for \protect\( r=\frac{K}{J_{1}}(K_{1}=K_{2}=K)\protect \)
\protect\( =0.4\protect \) \protect\( (a)\protect \), \protect\( 0.2\protect \)
\protect\( (b)\protect \) and \protect\( 0.0\protect \) \protect\( (c)\protect \).}
\end{figure}

\begin{figure}
{\centering \resizebox*{4in}{2in}{\includegraphics{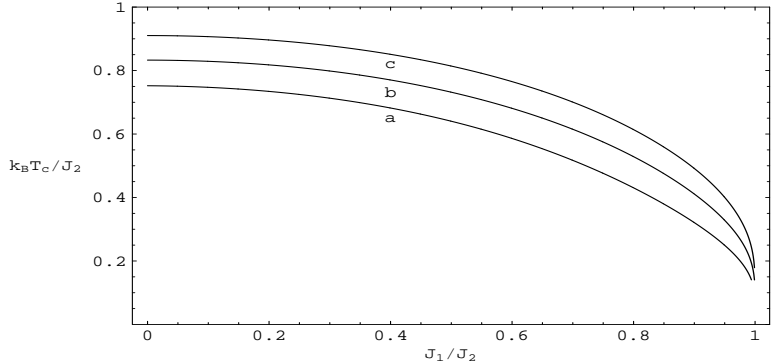}} \par}

\caption{A plot of \protect\( \frac{k_{B}T_{c}}{J_{2}}\protect \), where
\protect\( T_{c}\protect \) is the critical entanglement temperature,
versus \protect\( \frac{J_{1}}{J_{2}}\protect \), for \protect\( r=0.4\protect \)
\protect\( (a)\protect \), \protect\( 0.2\protect \) \protect\( (b)\protect \)
and \protect\( 0.0\protect \) \protect\( (c)\protect \).}
\end{figure}

\section*{III. Entanglement Witness}

We now consider the \( S=\frac{1}{2} \) polyoxovanadate AFM compound
\( V12 \) and show that the magnetic susceptibility \( \chi  \)
serves as an EW for this compound. The magnetic properties of this
system are well described by considering only the central square of
localized \( V^{4+} \)ions \cite{key-36}. These ions form a square
plaquette of \( S=\frac{1}{2} \) localized spins described by the
isotropic Heisenberg AFM Hamiltonian with only n.n. interactions \( (J_{2}=K_{1}=K_{2}=0 \)
in equation (2)). As shown in Ref. \cite{key-36}, the \( V12 \)
compound can be treated as a collection of independent \( S=\frac{1}{2} \)
tetramers with AFM n.n. interactions of strength \( \frac{J_{1}}{k_{B}}\simeq 17.6 \)
K. In fact, the theoretical expression for the magnetic susceptibility
\( \chi  \) of a tetramer gives a good fit (\( N \) independent
tetramers are to be considered in calculating \( \chi  \)) to the
experimental data for \( V12 \) (Fig. 2 of Ref. \cite{key-36}).
The susceptibility for a spin tetramer with only n.n. interaction
of strength \( J_{1} \) is given by 

\begin{equation}
\label{29}
\frac{\chi }{(g\mu _{B})^{2}/J_{1}}=\beta J_{1}\frac{10e^{-3\beta J_{1}}+4e^{-2\beta J_{1}}+2e^{-\beta J_{1}}}{1+e^{-2\beta J_{1}}+3e^{-\beta J_{1}}+6e^{-2\beta J_{1}}+5e^{-3\beta J_{1}}}
\end{equation}
Following Ref. \cite{key-28}, the magnetic susceptibility, \( \chi _{\alpha } \),
along the direction \( \alpha  \) \( (\alpha =x,y,z) \) can be written
as

\begin{equation}
\label{30}
\chi _{\alpha }=\frac{\left( g\mu _{B}\right) ^{2}}{k_{B}T}\left\langle \left( M_{\alpha }\right) ^{2}\right\rangle 
\end{equation}
where \( M_{\alpha }=\sum _{j}S^{\alpha }_{j} \) denotes the magnetization
along \( \alpha  \). The expression in (30) holds true when the external
magnetic field is zero and the Hamiltonian is isotropic in spin space.
The angular brackets in (30) denote the thermal expectation value.
The susceptibility \( \chi _{\alpha } \) can further be written as 

\begin{equation}
\label{31}
\chi _{\alpha }=\frac{\left( g\mu _{B}\right) ^{2}}{k_{B}T}\sum _{i,j}\left\langle S^{\alpha }_{i}S^{\alpha }_{j}\right\rangle 
\end{equation}
Due to the isotropy of the Hamiltonian, \( \chi _{x}=\chi _{y}=\chi _{z}=\chi  \)
and we can write

\begin{equation}
\label{32}
\chi =\frac{(g\mu _{B})^{2}}{k_{B}T}[\frac{N}{4}+\frac{2}{3}\sum _{i<j}\left\langle S_{i}.S_{j}\right\rangle ]
\end{equation}
where \( N \) is the total number of interacting spins. The summation
of expectation values in (32) can be considered as the expectation
value of the sum \( H_{S} \) of interaction terms describing all-to-all
spin couplings. The expectation value of \( H_{S} \) has an overall
negative contribution to \( \chi  \) because of AFM correlations.
\( H_{S} \) has the nature of a Hamiltonian and the maximum negative
expectation value is given by the ground state energy of \( H_{S} \).
For separable states, the energy minimum is given by the ground state
energy of the equivalent classical Hamiltonian \cite{key-32,key-33}.
For all-to-all spin couplings, the minimum energy separable state
is described by any spin configuration with total spin vector zero.
For \( N \)=4 (spin tetramer), the classical ground state is given
by the N\'{e}el state and \( \left\langle H_{S}\right\rangle  \)
= \( -\frac{1}{2} \). For general separable states, \( \left\langle H_{S}\right\rangle  \)
has a lesser negative contribution to \( \chi  \) and one can write
down the inequality 

{\centering \begin{equation}
\label{33}
\chi \geq \frac{(g\mu _{B})^{2}}{k_{B}T}\frac{2}{3}
\end{equation}
\par}

\noindent for separable, i.e., unentangled states. Figure 7 shows
a plot of \( \frac{\chi }{n(g\mu _{B})^{2}/J_{1}} \) versus \( T \)
(Curve \( a \) ) for \( n \) independent tetramers, the case of
\( V12 \). The expression for the susceptibility of a single tetramer
is given in (29). Curve \( b \) represents the \( \chi  \) versus
\( T \) curve describing the equality in (33). In plotting the curves,
the value of \( \frac{J_{1}}{k_{B}} \) is taken as \( 17.6 \) K,
the experimental estimate for \( V12 \). The intersection point of
the two curves provides an estimate, \( T_{c}\simeq 25.4 \) K, of
the critical entanglement temperature below which entanglement is
present in \( V12 \). The theoretical value of the critical temperature,
above which the two-spin entanglement disappears is obtained from
Eq. (26), with \( r_{1}=r_{2}=0 \), as \( T^{(1)}_{c}\simeq 15.2 \)
K. Since \( T_{c}>T_{c}^{(1)} \), only multipartite entanglement
is present in the thermal state of the tetramer for \( T_{c}^{(1)}<T<T_{c} \)
.
\begin{figure}
{\centering \resizebox*{4in}{2in}{\includegraphics{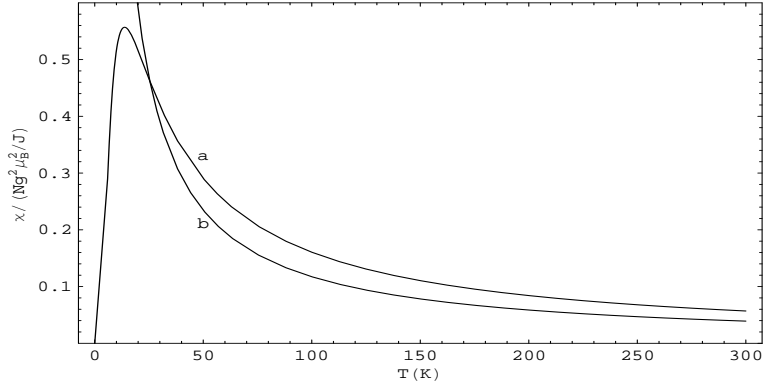}} \par}

\caption{A plot of \protect\( \frac{\chi }{n\left( g\mu _{B}\right) ^{2}/J_{1}}\protect \)
( dimensionless unit ) versus \protect\( T\protect \) (Curve \protect\( a\protect \)
) for \protect\( n\protect \) independent tetramers, as is the case
of \protect\( V12\protect \). Curve \protect\( b\protect \) represents
the \protect\( \chi \protect \) versus \protect\( T\protect \) curve
describing the equality in \protect\( (33)\protect \). The intersection
point of the two curves represents the critical entanglement temperature
\protect\( T_{c}\simeq 25.4\protect \) K with \protect\( \frac{J_{1}}{k_{B}}\simeq 17.6\protect \)
K. }
\end{figure}

We now examine whether four-spin entanglement exists in the thermal
state of the tetramer. This is done by calculating the state preparation
fidelity \( F \) defined as 

\begin{equation}
\label{34}
F(\rho )=\left\langle \psi _{GHZ}\right| \rho (T)\left| \psi _{GHZ}\right\rangle 
\end{equation}
where \( \left| \psi _{GHZ}\right\rangle =\frac{1}{\sqrt{2}}\left( \left| \uparrow \downarrow \uparrow \downarrow \right\rangle +\left| \downarrow \uparrow \downarrow \uparrow \right\rangle \right)  \)
is the four-spin Greenberger-Horne-Zeilinger (GHZ) state \cite{key-37}.
The sufficient condition for the four-particle (\( N=4 \) ) entanglement
is given by\begin{equation}
\label{35}
F(\rho )>\frac{1}{2}
\end{equation}
For a tetramer with only n.n. interactions of strength \( J_{1} \),
\( F(\rho ) \) is calculated as

\begin{equation}
\label{36}
F(\rho )=\frac{\frac{1}{3}e^{-\beta J_{1}}+\frac{2}{3}e^{2\beta J_{1}}}{5e^{-\beta J_{1}}+7+3e^{\beta J_{1}}+e^{2\beta J_{1}}}
\end{equation}
\( F(\rho )=\frac{2}{3} \),i.e., \( >\frac{1}{2} \) as \( T\rightarrow 0 \)
indicating the presence of four-spin entanglement in the ground state
of the tetramer. The critical entanglement temperature, \( T^{(4)}_{c} \),
beyond which the four-spin entanglement vanishes is obtained from
a solution of the equation \( F(\rho )=\frac{1}{2} \). The value
obtained is \( \frac{k_{B}T^{(4)}_{c}}{J_{1}}\simeq 0.417 \) from
which \( T^{(4)}_{c}\simeq 7.4 \) K, assuming \( \frac{J_{1}}{k_{B}}\simeq 17.6 \)
K as in the case of the compound \( V12 \). One finds that \( T^{(4)}_{c} \)
is less than the critical temperature \( T^{(1)}_{c}\simeq 15.2 \)
K. We next consider a tetramer with n.n., diagonal and four-spin exchange
interactions of strength \( J_{1},J_{2} \) and \( K_{1}=K_{2}=K \)
respectively. Fig. 8 shows a plot of \( \frac{k_{B}T^{(4)}_{c}}{J_{1}} \)
versus \( \frac{J_{2}}{J_{1}} \) for \( r=\frac{K}{J_{1}}=0.4 \)
\( (a), \) \( 0.2 \) \( (b) \) and \( 0.0 \) (c) respectively.
For a fixed value of \( \frac{J_{2}}{J_{1}} \), the critical temperature
for four-spin entanglement decreases as the strength of the four-spin
interaction increases.

\begin{figure}
{\centering \resizebox*{4in}{2in}{\includegraphics{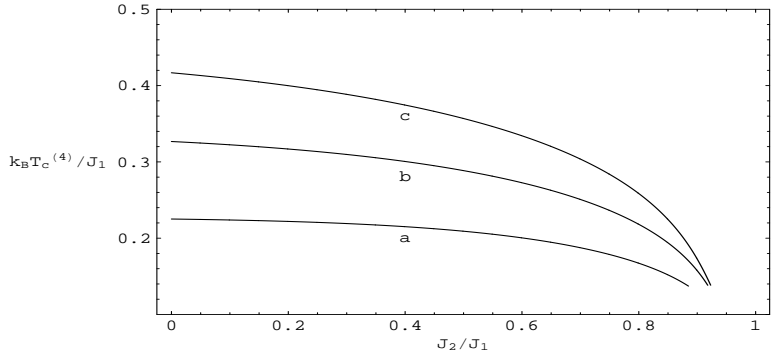}} \par}

\caption{Plot of \protect\( \frac{k_{B}T^{(4)}_{c}}{J_{1}}\protect \) versus
\protect\( \frac{J_{2}}{J_{1}}\protect \) for \protect\( r=\frac{K}{J_{1}}=0.4\protect \)
\protect\( (a)\protect \), \protect\( 0.2\protect \) \protect\( (b)\protect \)
and \protect\( 0.0\protect \) \protect\( (c)\protect \). \protect\( T^{(4)}_{c}\protect \)is
the critical entanglement temperature above which the four-spin entanglement
is zero.}
\end{figure}

The tetramer with \( J_{1}=J_{2}=J \) and \( K=0 \), alternatively
described as the tetrahedron, provides an interesting example of the
magnetic susceptibility \( \chi  \) serving as a witness for entanglement
other than the entanglement between individual spins. The two-spin
entanglement vanishes in the thermal state of the tetrahedron. The
same is true when \( T=0 \) and the system is at the QCP \( J_{1}=J_{2} \).
 Figure 9 shows the EW curves for \( \chi  \) (the same inequality
bound as in (33) holds true) which intersect at a finite temperature
\( \frac{k_{B}T_{c}}{J}\simeq 1.9 \) showing that the thermal states
are entangled below the critical temperature corresponding to the
intersection point. The entanglement is exclusively multipartite in
nature. Figure 10 shows the EW curves for \( \chi  \) in the case
of a symmetric trimer described by the \( S=\frac{1}{2} \) Heisenberg
AFM Hamiltonian 

\begin{equation}
\label{37}
H_{trimer}=J(S_{1}.S_{2}+S_{2}.S_{3}+S_{3}.S_{1})
\end{equation}
In this case, it is well-known \cite{key-37,key-44} that there is
no pairwise entanglement both at \( T=0 \) and at finite temperatures.
In the inequality for \( \chi  \) (Eq.(33)), the factor \( \frac{2}{3} \)
is replaced by the factor \( \frac{1}{2} \). In the classical ground
state of \( H_{S} \), the interacting spins form angles of \( \frac{2\pi }{3} \)
with each other. The critical temperature is given by \( \frac{k_{B}T_{c}}{J}\simeq 1.4 \)
. Again, only multipartite entanglement is present in the thermal
state of the symmetric trimer.

Another EW, which provides an estimate of critical entanglement temperature
\( T_{c} \), is based on energy \cite{key-32,key-33}. The entanglement
gap \( G_{E} \) is defined as 

\begin{equation}
\label{38}
G_{E}=E_{sep}-E_{0}
\end{equation}
where \( E_{0} \) is the ground state energy of the Hamiltonian \( H \)
describing the system and \( E_{sep} \) is the minimum energy of
the separable states. If \( G_{E} \) is > 0, a finite energy range
exists over which all states are entangled. For a positive entanglement
gap \( G_{E}>0 \), one can define an EW 

\begin{equation}
\label{39}
Z_{EW}=H-E_{sep}I
\end{equation}
where \( I \) represents the identity on the full Hilbert space.
For any separable state, \( Tr(Z_{EW}\rho _{sep)})\geq 0 \). If the
state is entangled, \( Tr(Z_{EW}\rho _{ent)}) \) is < 0. For example,
if the state belongs to the ground state manifold, \( Tr(Z_{EW}\rho _{0})=E_{0}-E_{sep}<0. \)
\( Z_{EW} \) thus acts as an EW.The entanglement gap temperature
\( T_{E} \) is given by \( U(T_{E})=E_{sep} \), where \( U(T) \)
 \( (=\left\langle H\right\rangle =-\frac{1}{Z}\frac{\partial Z}{\partial \beta }) \)
is the thermal energy at temperature \( T. \) For \( T<T_{E} \),
the thermal state is entangled and hence \( T_{E} \) is a measure
of the critical entanglement temperature. \( E_{sep} \) is given
by the ground state energy of the corresponding classical spin model
\cite{key-32,key-33}. For a square of spins, \( E_{sep}=-J_{1} \),
as the classical ground state is given by the N\'{e}el state. The
expression for \( U(T) \) is obtained from the partition function
\( Z \) of the square of spins As shown in \cite{key-33}, in the
case of bipartite graphs and lattices, the EW detects only bipartite
entanglement. Thus \( T_{E} \) for the square of spins has an identical
magnitude as that of \( T^{(1)}_{c} \) at and above which such entanglement
vanishes. In the case of non-bipartite graphs and lattices, the EW
can detect multipartite entanglement. The tetrahedron and the symmetric
trimer are examples of non-bipartite graphs. \( E_{sep} \) in these
two cases can readily be calculated as \( E_{sep}=-0.5J \) (tetrahedron)
and \( E_{sep}=-\frac{3}{8}J \) (symmetric trimer). The entanglement
temperature \( T_{E} \) has the magnitude \( \frac{k_{B}T_{E}}{J}\simeq 1.9 \)
(tetrahedron) and \( \frac{k_{B}T_{E}}{J}\simeq 1.4 \) (symmetric
trimer). In both the cases, two-spin entanglements are absent and
the entanglement present in the system for \( T<T_{E} \) is multipartite
in nature.

\begin{figure}
{\centering \resizebox*{4in}{2in}{\includegraphics{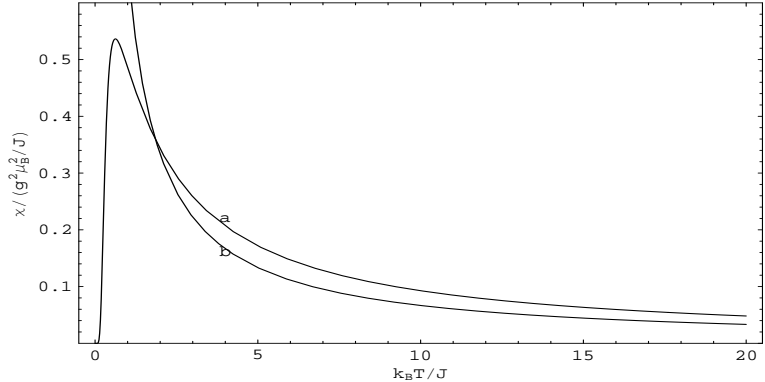}} \par}

\caption{The EW curves for \protect\( \chi \protect \) in the case of a symmetric
tetrahedron with \protect\( J_{1}=J_{2}=J\protect \) and \protect\( K=0\protect \).}
\end{figure}

\begin{figure}
{\centering \resizebox*{4in}{2in}{\includegraphics{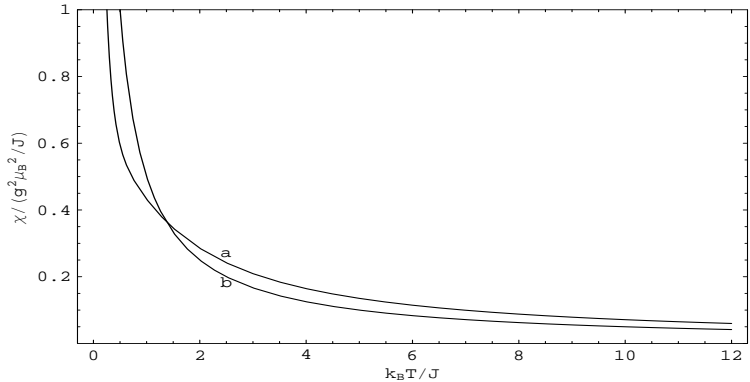}} \par}

\caption{The EW curves for \protect\( \chi \protect \) in the case of the
symmetric trimer.}
\end{figure}

\section*{IV. Summary and Discussion}

In this paper, we consider a spin tetramer \( (S=\frac{1}{2}) \)
with n.n., diagonal and four-spin AFM exchange interactions of strength
\( J_{1},J_{2} \) and \( K_{1}=K_{2}=K \) respectively. The significance
of the inclusion of three-spin and four-spin interactions in spin
Hamiltonians of interest has been pointed out earlier \cite{key-45,key-46}.
We study the ground state and thermal entanglement properties of the
tetramer in the various limiting cases. At \( T=0 \), QPTs occur
as the exchange interaction strengths are tuned to certain critical
values.We focus on a particular QPT at \( J_{1}=J_{2}=J \) \( (K<\frac{4}{5}J) \)
as the other QPTs exhibit similar features. The QPT point separates
two RVB ground states, \( \psi _{RVB1} \) and \( \psi _{RVB2} \).
The entanglement between two spins is determined by calculating the
concurrences \( C_{12},C_{23},C_{34},C_{41} \) and \( C_{13},C_{24} \).
The n.n. concurrences are non-zero only in \( \psi _{RVB1} \) and
the other two concurrences associated with diagonal spins are non-zero
only in \( \psi _{RVB2} \). The \( 1- \)tangle \( \tau _{1} \),
a measure of the global entanglement, has the value \( 1 \) in both
\( \psi _{RVB1} \) and \( \psi _{RVB2} \). The entanglement measure
\( \tau _{1}, \) defined for pure states, has to be generalized to
\( \tau _{1}^{min} \) at the QPT point where the ground state is
doubly degenerate. The entanglement ratio \( R=\frac{\tau _{2}}{\tau _{1}^{min}} \)
has the value zero at the transition point if \( \tau _{1}^{min}\neq 0 \)
and is undefined otherwise. Away from the transition point, \( R=0.5 \)
in the ground state \( \psi _{RVB1} \) and \( 1.0 \) in the ground
state \( \psi _{RVB2} \). A better evidence of the first order QPT
is provided by the jumps in both the n.n. and diagonal concurrences
\cite{key-18,key-23}.

The study of finite temperature entanglement properties again shows
the existence of two distinctive parameter regimes. The n.n. concurrences
are non-zero only when \( J_{2}<J_{1}(K<\frac{4J_{1}}{5}) \) and
the concurrences associated with diagonal spins are non-zero only
when \( J_{1}<J_{2}(K<\frac{4J_{2}}{5}) \). At \( J_{1}=J_{2} \),
all the six concurrences are zero. The critical entanglement temperature,
\( T_{c} \), beyond which entanglement between two spins disappears,
is computed. The magnitude of \( T_{c} \) is highest when \( J_{2}=0 \)
and \( K=0 \). For fixed values of \( J_{1} \) and \( J_{2} \),
\( T_{c} \) decreases as the strength of the four-spin interaction
increases. A measure of the four-spin entanglement in the thermal
state of the tetramer is obtained by calculating the fidelity \( F(\rho ) \).
The critical temperature, \( T^{(4)}_{c} \), beyond which the four-spin
entanglement disappears is calculated and one finds that at fixed
values of \( J_{1} \) and \( J_{2} \), the magnitude of \( T^{(4)}_{c} \)
decreases as the strength \( K \) of the four-spin interaction increases. 

Molecular or nanomagnets provide examples of spin systems in which
the dominant exchange interactions are confined to small spin clusters
like dimers, trimers and tetramers. In several cases, the magnetic
properties can be well explained by treating the solid to consist
of independent spin clusters. We consider one such compound, \( V12 \),
which is a collection of spin tetramers with only n.n. exchange interactions.
Treating the magnetic susceptibility \( \chi  \) as an EW, the critical
temperature, \( T_{c} \), below which entanglement is present in
the system, is estimated from the experimental data on \( \chi  \).
The entanglement includes both bipartite and multipartite entanglement
with \( T_{c} \) \( \simeq  \) 25.4 K in the case of \( V12 \)
\( (\frac{J}{k_{B}}\simeq 17.6 \) K). From theoretical calculations,
the critical temperature \( T_{c}^{(1)} \), beyond which bipartite
entanglement vanishes is given by \( T^{(1)}_{c} \)\( \simeq 15.2 \)
K. Since \( T_{c}^{(1)}<T_{c} \), multipartite entanglement in the
system persists upto a higher temperature. The entanglement contents
of the thermal states of the tetrahedron and the symmetric trimer
are shown to be exclusively multipartite in nature. An EW based on
energy provides evidence of bipartite entanglement in the case of
a square of spins (relevant for \( V12 \)) and multipartite entanglement
in the cases of the tetrahedron and the symmetric trimer. The EW based
on susceptibility \( \chi  \) can detect both bipartite and multipartite
entanglement. The EW based on energy detects only bipartite entanglement
when the spin system is defined on a bipartite graph or lattice. The
latter EW can detect multipartite entanglement only in the case of
a non-bipartite graph or lattice. The critical entanglement temperature
\( T_{c} \) and the entanglement gap temperature \( T_{E} \) have
identical values if the corresponding EWs both detect entanglement. 

After our work was completed, we learnt of a new inequality for susceptibility
serving as an EW \cite{key-51,key-52}. The inequality can be derived
using the sum uncertainty relation for spin\( -\frac{1}{2} \) operators
\cite{key-51,key-53}. When \( \chi _{x}=\chi _{y}=\chi _{z}=\chi  \),
the separability criterion for a single cluster of \( N \) spins
is given by

\begin{equation}
\label{40}
\chi \geq \frac{(g\mu _{B})^{2}}{k_{B}T}\frac{N}{6}
\end{equation}
The results reported by us, using slightly different arguments, are
special cases of the general condition (40) for \( N \) = 3 and 4.
We can generalize our derivation in the following manner to obtain
(40). We start with the identity

\begin{equation}
\label{41}
H_{S}=\sum _{i<j}S_{i}S_{j}=\frac{1}{2}(S^{2}-\sum ^{N}_{i=1}S_{i}^{2})
\end{equation}
where \textbf{\emph{S}} is the total spin vector. The maximum negative
contribution to \( \chi  \) is obtained for \textbf{\emph{S}} =0.
Thus for separable states with \( \left\langle S_{i}^{2}\right\rangle  \)
= \( \frac{1}{4} \), Eq.(32) reduces to the inequality in (40). There
is now a wealth of experimental data on molecular magnets and other
magnetic systems which are yet to be analyzed in terms of the entanglement
properties of the systems \cite{key-35,key-50}. Appropriate finite
temperature measures of the different types of entanglement need to
be developed so that contact between theory and experiments can be
made. A challenging task ahead is to develop suitable EWs which provide
signatures of the different types of entanglement in the experimental
data.\\
\textbf{Acknowledgment}. Amit Tribedi is supported by the Council
of Scientific and Industrial Research, India under Grant No. 9/15
(306)/ 2004-EMR-I.


\begin{thebibliography}{10}
\bibitem{key-1}M. A. Nielsen and I. L. Chuang, Quantum Computation and Quantum Information
(Cambridge University Press, Cambridge, 2000)
\bibitem{key-2}D. Loss and D. P. DiVincenzo, Phys. Rev. A 57, 120 (1998); G. Burkard,
D. Loss and D. P. DiVincenzo, Phys. Rev. B 59, 2070 (1999)
\bibitem{key-3}B. E. Kane, Nature 393, 133 (1998)
\bibitem{key-4}D. P. DiVincenzo, D. Bacon, J. Kempe, G. Burkard and K. B. Whaley,
Nature (London) 408, 339 (2000)
\bibitem{key-5}J. J. Garcia-Ripoll, M. A. Martin-Delgado and J. I. Cirac, Phys. Rev.
Lett. 93, 250405 (2004)
\bibitem{key-6}M. Popp, F. Verstraete, M. A. Martin-Delgado and J. I. Cirac, quant-ph/0411123
\bibitem{key-7}K. M. O'Connor and W. K. Wootters, Phys. Rev. A 63, 052302 (2001);
W. K. Wootters, Phys. Rev. Lett. 80, 2245 (1998)
\bibitem{key-8}M.C . Arnsen, S. Bose ad V. Vedral, Phys. Rev. Lett. 87, 017901 (2001);
D. Gunlycke et al., Phys. Rev. A 64, 042302 (2001)
\bibitem{key-9}M. A. Nielsen, Ph. D. Thesis, University of Mexico, 1998, quant-ph/0011036
\bibitem{key-10}X. Wang, Phys. Rev. A 64, 012313 (2001); Phys. Lett. A 281, 101 (2001)
\bibitem{key-11}X. Wang, Phys. Rev. A 66, 034302 (2002)
\bibitem{key-12}D. Bru\( \beta  \), N. Dutta, A. Ekert, L. C. Kwek and C. Macchiavello,
quant-ph/0411080
\bibitem{key-13}U. Glasser, H. B\"{u}ttner and H. Fehske, Phys. Rev. A 68, 032318
(2003)
\bibitem{key-14}T.-C. Wei, D. Das, S. Mukhopadhyay, S. Vishveshwara and P. M. Goldbart,
quant-ph/0405162
\bibitem{key-15}A. Osterloh, L. Amico, G. Falci and R. Fazio, Nature (London) 416,
608 (2002)
\bibitem{key-16}T. J. Osborne ad M. A. Nielsen, Phys. Rev. A 66, 032110 (2002)
\bibitem{key-17}S. Sachdev, Quantum Phase Transitions (Cambridge University Press,
Cambridge, 1999)
\bibitem{key-18}L.-A. Wu, M. S. Sarandy and D. A. Lidar, Phys. Rev. Lett. 93, 250404
(2004)
\bibitem{key-19}J. Vidal, G. Palacios and R. Mosseri, Phys. Rev. A 69, 022107 (2004)
\bibitem{key-20}G. Vidal, J. I. Latorre, E. Rico and A. Kitaev, Phys. Rev. Lett. 90,
227902 (2003)
\bibitem{key-21}F. Verstraete, M. Popp and J. I. Cirac, Phys. Rev. Lett. 92, 027901
(2004)
\bibitem{key-22}R. Somma, G. Ortiz, H. Barnum, E. Knill and L. Viola, Phys. Rev. A
70, 042311 (2004)
\bibitem{key-23}I. Bose and E. Chattopadhyay, Phys. Rev. A 66, 062320 (2002)
\bibitem{key-24}F. C. Alcaraz, A. Saguia and M. S. Sarandy, Phys. Rev. A70, 032333
(2004)
\bibitem{key-25}J. Vidal, R. Mosseri and J. Dukelsky, Phys. Rev. A 69, 054101 (2004)
\bibitem{key-26}X.-F.Qian, T. Shi, Y. Li, Z. Song and C.-P. Sun, quant-ph/0502121
\bibitem{key-27}S. Ghosh, T. F. Rosenbaum, G. Aeppli and S. N. Coppersmith, Nature
425, 48 (2003); V. Vedral, Nature 425, 28 (2003)
\bibitem{key-28}\u{C}. Brukner, V. Vedral and A. Zeilinger, quant-ph/0410138
\bibitem{key-29}X. Wang and P. Zanardi, Phys. Lett. A 301, 1 (2002)
\bibitem{key-30}V. Vedral, New J. Phys. 6, 22 (2004)
\bibitem{key-31}\u{C}. Brukner and V. Vedral, quant-ph/0406040
\bibitem{key-32}G. T\'{o}th, quant-ph/0406061
\bibitem{key-33}M. R. Dowling, A. C. Doherty and S. D. Bartlett, quant-ph/0408086
\bibitem{key-34}L.-A. Wu, S. Bandyopadhyay, M. S. Sarandy and W. A. Lidar, quant-ph/0412099
\bibitem{key-35}J. T. Haraldsen, T. Barnes and J. L. Musfeldt, Phys. Rev. B 71, 064403
(2005)
\bibitem{key-36}D. Procissi, A. Shastri, I. Rousochatzakis, M. AI Rifafi, P. K\"{o}gerler
and M. Luban, Phys. Rev. B 69, 094436 (2004)
\bibitem{key-37}X. Wang, Phys. Rev. A 66, 044305 (2002)
\bibitem{key-38}S.- J. Gu, H. Li, Y.- Q. Li and H. -Q. Lin, quant-ph/0403026
\bibitem{key-39}V. Coffman, J. Kundu and W. K. Wootters, Phys. Rev. A 61, 052306 (2000)
\bibitem{key-40}T. Roscilde, P. Verrucchi, A. Fubini, S. Haas and V. Tognetti, Phys.
Rev. Lett. 93, 167203 (2004)
\bibitem{key-48}T. Roscilde, P. Verrucchi, A. Fubini, S. Haas and V. Tognetti, quant-ph/0412098
\bibitem{key-43}H. Fu, A. I. Solomon and X. Wang, quant-ph/0401015
\bibitem{key-44}X. Wang , H. Fu and A. I. Solomon, J. Phys. A 34, 11307 (2001)
\bibitem{key-45}J. K. Pachos and M. B. Plenio, quant-ph/0401106
\bibitem{key-46}A. Mizel and D. A. Lidar, quant-ph/0401081
\bibitem{key-50}M. U.- Kartin, S.-J. Hwu and J. A. Clayhold, Inorganic Chemistry 42,
2405 (2003)
\bibitem{key-51}M. Wie\'{s}niak, V. Vedral and \u{C}. Brukner, quant-ph/0503037
\bibitem{key-52}T. V\'{e}rtesi and E. Bene, cond-mat/0503726
\bibitem{key-53}H. F. Hofmann and S. Takeuchi, Phys. Rev. A 68, 032103 (2003)
\end{thebibliography}
\end{document}